\documentclass[twocolumn,eqsecnum,showpacs,showkeys,amsmath,amssymb,nofootinbib,superscriptaddress,floatfix]{revtex4}

\usepackage{graphicx}
\usepackage{subfigure}
\usepackage{bm}

\makeatletter
\newcommand\erfc{\mathop{\operator@font erfc}\nolimits}
\def\slashchar#1{\setbox0=\hbox{$#1$}
   \dimen0=\wd0 \setbox1=\hbox{/} \dimen1=\wd1
   \ifdim\dimen0>\dimen1 \rlap{\hbox to \dimen0{\hfil/\hfil}} #1
   \else  \rlap{\hbox to \dimen1{\hfil$#1$\hfil}} / \fi}

\makeatother

\begin{document}
\title{Rapid hydrodynamic expansion  in
 relativistic heavy-ion collisions\footnote{Supported by 
Polish Ministry of Science and Higher Education under
grant N202~034~32/0918}}
\author{Piotr Bo\.zek}
\email{Piotr.Bozek@ifj.edu.pl}
\affiliation{The H. Niewodnicza\'nski Institute of Nuclear Physics,
PL-31342 Krak\'ow, Poland} \affiliation{
Institute of Physics, Rzesz\'ow University, 
PL-35959 Rzesz\'ow, Poland}
\author{Iwona Wyskiel}
\affiliation{The H. Niewodnicza\'nski Institute of Nuclear Physics,
PL-31342 Krak\'ow, Poland}
\date{\today}

\begin{abstract}
Hydrodynamic expansion  of the hot fireball created in
 relativistic Au-Au collisions at $\sqrt{s}=200$GeV in $3+1$-dimensions 
is  studied.
We obtain a  
simultaneous, satisfactory 
 description 
of the transverse momentum spectra, 
elliptic flow and pion correlation radii for 
different collision centralities and different rapidities.
Early initial time of the evolution is required to
 reproduce the interferometry data, which provides a strong 
indication of the early onset of collectivity. We can also 
 constraint the shape of the initial energy density in the 
beam direction, with 
 a relatively high initial energy density at the  center of the fireball.
\end{abstract}



\maketitle

\section{Introduction}

A multitude of experimental data  from the Relativistic Heavy Ion 
Collider
 (RHIC)
indicate that  dense, collectively expanding matter is 
created in ultrarelativistic nuclear collisions
\cite{Arsene:2004fa,Back:2004je,Adams:2005dq,Adcox:2004mh}. Ratios 
of multiplicities of
different particles produced in central collisions can be described assuming 
chemical equilibration of  particle abundances 
\cite{BraunMunzinger:2001ip,Andronic:2005yp,Cleymans:2004pp,Florkowski:2001fp,Rafelski:2004dp,Becattini:2005xt}.
 Transverse momentum spectra of particles 
produced at central rapidities are thermal up to transverse 
momenta of about $2$GeV/c.
Particle spectra result from  a collective, transverse expansion 
of the matter
 coupled with subsequent thermal emission 
\cite{Schnedermann:1993ws,Kolb:2000sd}.  
Possible rescattering and resonance decays 
have  been modelled and  the conclusion, that at some 
stage of the expansion a dense locally equilibrated fireball is 
formed, remains unchanged \cite{Bass:2000ib,Nonaka:2006yn}.
Another measured quantity directly resulting from the collective expansion 
is the elliptic flow. For non-zero impact parameters the fireball is 
azimuthally asymmetric in the transverse plane, and its expansion imprints
the momentum distribution of final hadrons with measurable azimuthal asymmetry
\cite{Ollitrault:1992bk,Voloshin:2008dg,Teaney:2000cw,Huovinen:2001cy}.
The existence of the dense matter is demonstrated in yet another way by the
observation of the attenuation of the production of high energy hadrons 
in nuclear collisions. This effect is due to the energy loss of energetic 
partons while traversing the dense fireball \cite{Back:2003ns,Adler:2003ii,Adams:2003im,Arsene:2003yk,Gyulassy:1990ye,Baier:2000mf}.

Relativistic hydrodynamics is very well suited for the description of the 
 collective phase of the fireball expansion \cite{Teaney:2000cw,Hirano:2001yi,Kolb:2003dz,Hirano:2002ds,Huovinen:2003fa,Hama:2005dz,Huovinen:2006jp,Hirano:2005xf,Hirano:2007xd,Nonaka:2007nn,Nonaka:2006yn,Broniowski:2008vp,Hirano:2008hy,Akkelin:2008eh,Andrade:2008xh}. Assuming 
local thermal equilibration,
perfect fluid hydrodynamics can be used. Starting from an initial
 energy density profile, the fluid expands and cools down. In the 
process, gradients of the pressure cause the acceleration of the fluid elements
and  collective flow velocity is formed. 
In the longitudinal (beam)  direction  Bjorken  flow with velocity $v_z=z/t$
is usually assumed in the initial conditions. On the other hand, the appearance
of a substantial transverse flow  can be considered as a robust signature of
 the formation of strongly interacting matter in the overlap region 
of heavy-ion collisions. Most of the hydrodynamic calculations 
modeling nuclear collisions 
at RHIC energies assume boost-invariance \cite{Bjorken:1983}
in the beam direction. Such approaches 
are effectively $2+1$-dimensional ($2+1$D) and are restricted to  
 central rapidities. Existing experimental data outside of the central
 rapidity region on  particle multiplicity and elliptic 
flow show that at RHIC energies the Bjorken boost-invariance 
 is not realized.
 Calculations exist for the general $3+1$D 
geometry of the collision
 \cite{Hirano:2001yi,Hirano:2002ds,Hama:2005dz,Nonaka:2006yn,Andrade:2008xh}. 
 They show that relativistic hydrodynamics can be 
applied  for a broad range of rapidities in central 
and semiperipheral collisions.  These studies  can describe 
 transverse momentum spectra and the elliptic flow of produced particles. 
On the other hand,
Hanbury Brown-Twiss  (HBT) correlations between identical particles cannot be 
accounted for
\cite{Hirano:2002ds,Aguiar:2001ac,Morita:2003mj,Morita:2006zz}.
In the present paper we investigate $3+1$D hydrodynamic expansion 
of the fireball and show that a simultaneous and  satisfactory 
description of the particle spectra and elliptic flow (for a broad range 
of rapidities) as well as of the HBT radii for central rapidities can be 
achieved.
The key ingredients of the model leading to this success are the use of a 
realistic equation of state without a first order phase transition and 
a relatively early start up time for the collective expansion.
This  hard equation of state and the small initial time
indicate that the initial state is a highly compressed 
matter with energy density of up to $100$GeV/fm$^{3}$ in central collisions.
In this work we use perfect fluid hydrodynamics. 
Shear viscosity or hadronic dissipative 
effects are known to modify final observables 
\cite{Muronga:2001zk,Muronga:2006zw,Baier:2006gy,Romatschke:2007mq,Chaudhuri:2006jd,Song:2007fn,Hirano:2007xd,Hirano:2005xf}, especially the elliptic flow.
The influence of viscosity effects on particle spectra or HBT radii is more 
difficult to be explicitly demonstrated, since such effects can be 
compensated by a change in the unknown initial time or energy density profile. 

\section{Hydrodynamic equations and initial conditions}
\label{sec:ini}

In a perfect fluid each element is locally in thermal equilibrium. At each
 point the fluid is characterized by its four velocity $u^\mu$, 
the energy density $\epsilon$, and the pressure $p$. The energy 
momentum tensor is
\begin{equation}
T^{\mu\nu}=(\epsilon+p)u^\mu u^\nu-g^{\mu\nu}p \ .
\end{equation}
Hydrodynamic equations
\begin{equation}
\partial_\mu T^{\mu\nu}=0 
\end{equation}
in the full 3+1D geometry represent 4 independent equations,
and together with the equation of state
 allow to calculate the evolution of the densities 
and velocities of the fluid starting from some initial conditions.
We use a realistic equation of state interpolating between
 lattice data at high temperature (above the critical temperature $T_c=170$MeV)
 and an equation of state of a noninteracting gas of massive 
hadrons at lower temperatures \cite{Chojnacki:2007jc}. This equation of state 
presents only a very moderate softening around the critical point. 
The use of this realistic equation of state is the key to the success 
of $2+1$D hydrodynamic description of RHIC data on
 transverse momentum spectra, elliptic flow and HBT radii
 \cite{Chojnacki:2007rq,Broniowski:2008vp}.
For the modelling of the expansion of the fireball created in 
ultrarelativistic collisions it is useful to define the proper 
time and space-time rapidity variables
\begin{equation}
\tau=\sqrt{t^2-z^2}\ , \ \ \
\eta=\frac{1}{2}\log\left(\frac{t+z}{t-z}\right) \ ,
\end{equation}
with $z$ the beam axis coordinate.
The four velocity is parameterized using the two components of 
the transverse velocity $u_x$ and $u_y$ and the longitudinal fluid rapidity $Y$
\begin{equation}
u^\mu=(\gamma \cosh Y, u_x, u_y, \gamma\sinh Y) \ ,
\end{equation}
where $\gamma=\sqrt{1+u_x^2+u_y^2}$. Hydrodynamic equations
 relate  four unknown functions~:
 the velocity fields $Y$, $u_x$, and $u_y$ and either the energy or the 
pressure.
In practice the numerical solution is more stable if instead 
of the energy density (pressure) 
 the logarithm of the temperature is used 
\begin{equation}
{\cal F}=\log(T/T_L) \ ,
\end{equation}
with $T_L$ a constant temperature. The velocities and ${\cal F}$ are 
function of $\tau, \eta, x, y$ and the hydrodynamic equations can be written in the following form
\begin{eqnarray}
&&D {\cal F} =- c_s^ 2\Big[ \gamma \left(\sinh(Y-\eta)\partial_\tau+\frac{\cosh(Y-\eta)}{\tau}\partial_\eta\right) Y \nonumber \\
& &  \left(\cosh(Y-\eta)\partial_\tau+\frac{\sinh(Y-\eta)}{\tau}\partial_\eta\right)\gamma \nonumber \\
& &+  \partial_x u_x +\partial_y u_y \Big] \nonumber \\
& & D u_x  =  -(1+u_x^2)\partial_x  {\cal F} - u_xu_y \partial_y {\cal F}\nonumber \\
& &- u_x \gamma \left(\cosh(Y-\eta)\partial_\tau+\frac{\sinh(Y-\eta)}{\tau}\partial_\eta\right){\cal F} \nonumber \\
& & D u_y  =  -(1+u_y^2)\partial_y  {\cal F} - u_xu_y \partial_x {\cal F}\nonumber \\
&& - u_y \gamma \left(\cosh(Y-\eta)\partial_\tau+\frac{\sinh(Y-\eta)}{\tau}\partial_\eta\right){\cal F} \nonumber \\
& &D (\gamma \sinh Y) =-\cosh Y\left[ \sinh(Y-\eta)\partial_\tau+
\frac{\cosh(Y-\eta)}{\tau}\partial_\eta \right. \nonumber \\
&& + \left. (u_x^2+u_y^2)(\cosh(Y-\eta)\partial_\tau+
\frac{\sinh(Y-\eta)}{\tau}\partial_\eta) \right] {\cal F} \nonumber \\
&&- \gamma\sinh Y \left( u_x\partial_x +u_y\partial_y\right) {\cal F} \ ,
\label{eq:hydro4}
\end{eqnarray} 
 where $c_s$ is the sound velocity  and
\begin{eqnarray}
D& =& u^\mu\partial_\mu = u_x\partial_x+u_y\partial_y\nonumber \\
&+&\gamma (\cosh(Y-\eta)\partial_\tau+
\frac{\sinh(Y-\eta)}{\tau}\partial_\eta) \ .
\end{eqnarray}
The first of Eqs (\ref{eq:hydro4}) is the entropy conservation
 equation
$\partial_\mu(u^\mu  s )=0$.

 The differential equations (\ref{eq:hydro4}) are solved as an evolution 
in proper time starting from some initial conditions at $\tau=\tau_0$. 
At the  initial time there is no transverse flow ($u_x=0$ and $u_y=0$), 
the initial longitudinal rapidity  follows the Bjorken  scaling flow
$Y(\tau_0,\eta,x,y)=\eta$.
Early initial time of the hydrodynamic evolution $\tau_0=0.25$fm/c implies
 a high energy density in the initial state. The system evolves for a 
longer time, which leads to a stronger transverse as well as longitudinal flow.
Experimental observation  of a strong rapidity dependence 
of the elliptic flow \cite{Back:2004mh} 
and of the particle densities \cite{Bearden:2004yx} indicates that the 
Bjorken scaling scenario is not realized at RHIC energies. It means that 
a Bjorken scaling plateau in the initial energy density distribution
 in space-time rapidity cannot extend over a large interval.
 In the transverse plane the energy
 density is assumed to be proportional to a combination of Glauber 
Model densities
 of wounded nucleons and  binary collisions.
The initial energy density distribution at impact parameter $b$ is
\begin{eqnarray}
\epsilon(\tau_0)&=&k  f(\eta-\eta_{sh})\left[
 \left( N_A(x,y)+ N_B(x,y)\right)
(1-\alpha) \right.\nonumber \\
&+& \left. 2 \alpha N_{bin}(x,y)\right] \ .
\label{eq:nwdens}
\end{eqnarray}
$N_A$ and $N_B$ are the densities of wounded nucleons from the right and
 left moving
nuclei respectively, $N_{bin}$ is the density of binary collisions
\begin{eqnarray}
N_A(x,y)& =& T(x-b/2,y)\left(1-\exp(-\sigma T(x+b/2,y)/A)\right) \nonumber \\
N_B(x,y)& =& T(x+b/2,y)\left(1-\exp(-\sigma T(x-b/2,y)/A)\right) \nonumber \\
N_{bin}(x,y)&=&\sigma T(x-b/2,y)T(x+b/2,y) 
\end{eqnarray}
and
\begin{equation}
T(x,y) =\int dz \rho(x,y,z) 
\end{equation}
is the thickness function calculated from the 
Woods-Saxon density of colliding nuclei
\begin{equation}
\rho(x,y,z)=\frac{\rho_0}{1+\exp\left((\sqrt{x^2+y^2+z^2}-R_A)/a\right)} \ .
\end{equation}
For Au nuclei ($A=197$) 
we take $\rho_0=0.17 \mbox{fm}^{-3}$, $R_A=6.38$fm and
$a=0.535$fm; the inelastic cross section is $\sigma=42$mb.
The density of wounded nucleons $N_A+N_B$ is used to calculate the
 total number of participants at each impact parameter, these numbers
 are used to fix the impact parameters corresponding to  centrality bins 
used in the analysis of the experimental data. 
The profile in the longitudinal direction is 
\begin{equation}
f(\eta)=\exp\left(-\frac{(\eta-\eta_0)^2}{2\sigma_\eta^2}\theta(|\eta|-\eta_0)
\right)
\label{eq:etaprofile}
\end{equation}
with a plateau of width $2\eta_0=2.0$
 units in space-time rapidity, and Gaussian tails
with half width $\sigma_\eta=1.3$. At each point in the transverse plane  the 
distribution in space-time
 rapidity is shifted by the center of mass rapidity of the
local fluid \cite{Hirano:2002ds}
\begin{equation}
\eta_{sh}=\frac{1}{2}\log\left(\frac{N_A+N_B+v_N(N_A-N_B)}
{N_A+N_B-v_N(N_A-N_B)}\right) \ ,
\label{eq:etash}
\end{equation}
where $v_N$ is the velocity of the projectile in the center of mass frame.
The coefficient of $k$ in Eq. (\ref{eq:nwdens}) is taken so that the 
energy density at the center of the fireball at zero impact parameter 
is $107$GeV/fm$^{3}$ at $\tau_0=0.25$fm/c. It corresponds to a temperature 
of $510$MeV, well above the critical temperature. The initial distributions
for other impact parameters are obtained from geometrical scaling (\ref{eq:nwdens})
only, with
 a contribution of binary collisions $\alpha=0.145$ 
\cite{Back:2004dy}.  This provides a satisfactory description of charged 
particle multiplicities
for centralities $0-40\%$.

\section{Evolution of the hot matter}
\label{sec:bulk}
\begin{figure}
\includegraphics[width=.3\textwidth]{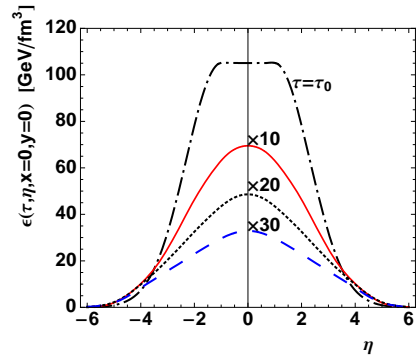}
\caption{(Color online) Energy density as  function 
of space-time rapidity for different proper times 
$\tau=0.25,\ 2, \ 4,\ 6$fm/c (dashed-doted, solid,
 dotted and dashed lines respectively).}
\label{fig:endens}
\end{figure}

\begin{figure}
\includegraphics[width=.3\textwidth]{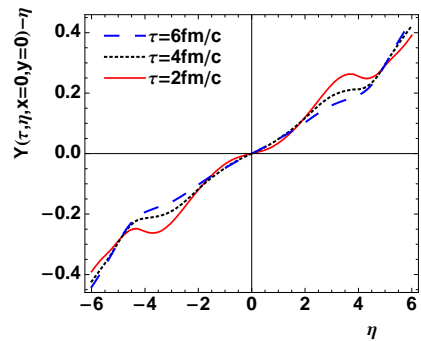}
\caption{(Color online) Deviation of the longitudinal 
fluid rapidity from the Bjorken flow $Y(\tau,\eta,x,y)-\eta$ 
for different proper times 
$\tau= 2, \ 4,\ 6$fm/c  ( solid,
 dotted and dashed lines respectively).}
\label{fig:rapflow}
\end{figure}

 The numerical solution of Eqs. (\ref{eq:hydro4}) is obtained as an 
evolution in proper time from initial densities (\ref{eq:nwdens}). 
The total entropy ($s$ is the entropy density)
\begin{equation}
S=\int  \gamma \cosh(Y-\eta) s \ dx dx d\eta
\end{equation}
is conserved to the accuracy of less than $0.5\%$.
At the very beginning of the evolution a very rapid longitudinal expansion 
occurs (Fig. \ref{fig:endens}).
The matter at the
edges of the plateau of the energy density distribution is  subject to 
longitudinal acceleration and eventually the 
 distribution becomes approximately a Gaussian,
 that grows wider in time. The longitudinal 
 acceleration is known to depend on the equation of state 
\cite{Satarov:2006iw}
and on  possible viscosity effects \cite{Bozek:2007qt}. For the perfect fluid
and a hard equation of state  the longitudinal expansion and 
acceleration is significant \cite{Satarov:2006iw}.

\begin{figure}
\includegraphics[width=.3\textwidth]{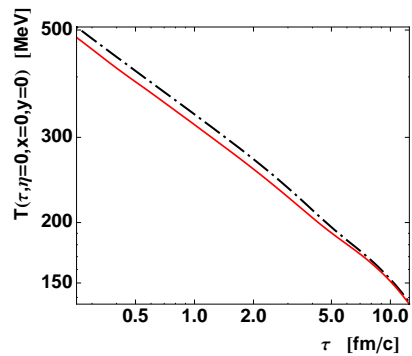}
\caption{(Color online) Temperature at the center of the fireball as 
function of the proper time from the $3+1$D (dashed-dotted line)
 and from the $2+1$D (solid line) evolutions ($b=2.1$fm).}
\label{fig:tempcool}
\end{figure}

For comparison we calculate a hydrodynamic evolution of the 
system assuming a $2+1$D boost invariant expansion, with the 
energy density profile in the transverse plane given by equation 
(\ref{eq:nwdens}), but without  $\eta$ dependence. 
The temperature at the  center
 of the fireball ($T=485$MeV for $b=0$) 
that reproduces the observed spectra is slightly lower than in the 
$3+1$D case. This is the effect of the additional cooling 
in the non-boost invariant geometry due to the longitudinal acceleration.
In $3+1$D the 
  longitudinal fluid rapidity is larger than in the Bjorken scaling solution 
(Fig. \ref{fig:rapflow}). At the center of the fireball the temperature 
drops down following the Bjorken formula $T\propto \tau^{-c_s^2}$ 
(Fig. \ref{fig:tempcool}) up to $\tau=2-3$fm/c. Later cooling from
 the transverse expansion  and in the case of $3+1$D additional
 longitudinal colling set in. As a result the life-time of the $2+1$D and 
$3+1$D systems is very similar, in spite of the fact that in the
 later case  the initial energy density is a factor $1.25$ higher.


\begin{figure}
\includegraphics[width=.3\textwidth]{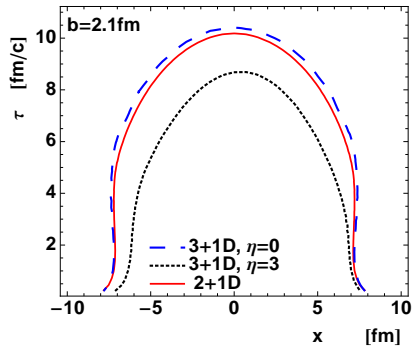}
\caption{(Color online) Freeze-out hypersurface $T_f=150$MeV,
for the impact parameter 
$b=2.1$fm in the  plane $(t-x),\ y=0$,
for $\eta=0$ (dashed line) and  $\eta=3$ (dotted line). The solid line 
represents the freeze-out hypersurface for the $2+1$D evolution. }
\label{fig:freeze1}
\end{figure}

\begin{figure}
\includegraphics[width=.3\textwidth]{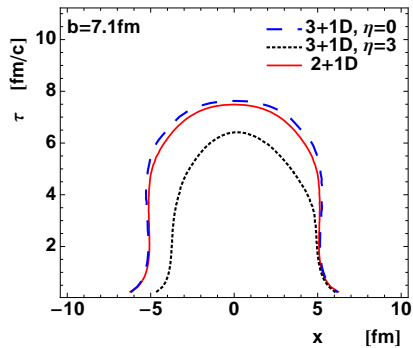}
\caption{(Color online) Same as Fig. \ref{fig:freeze1} but for $b=7.1$fm.}
\label{fig:freeze2}
\end{figure}

The hydrodynamic evolution is followed until freeze-out, that
 is assumed to happen at  fixed temperature, with particles
emitted from the freeze-out hypersurface without further rescattering. 
In the following, we present results for  two different freeze-out temperatures
 $T_f=150$ and $165$MeV.  
The freeze-out hypersurface is 
a three-dimensional surface in $\tau, \eta, x, y$ 
coordinates. Its shape can be deformed due to a strong collective
 flow and is deformed in the $\eta-x$ direction due to the shift 
in the space-time rapidity in the initial conditions (\ref{eq:etash}).
In Figs \ref{fig:freeze1} and \ref{fig:freeze2} is shown a cut ($\tau-x$) 
through the 
freeze-out hypersurface at the freeze-out temperature $T_f=150$MeV.
At the central space-time rapidity $\eta=0$ the freeze-out hypersurfaces in the
 $3+1$D and $2+1$D calculations are very similar. For central 
collisions the dense system exists for $10$fm/c. This short life-time 
of the fireball results in values of extracted HBT radii compatible with the
experiment. For large space-time rapidities the transverse size and 
the life-time of the fireball is smaller. 
The asymmetric shape of the freeze-out hypersurface 
for $\eta\neq 0$ is the result of the tilt in the initial flow
 of the matter (Eq. \ref{eq:etash}). 

Most general freeze-out hypersurfaces realized in 
the  hydrodynamic expansion can be parameterize using $3$ angles
\begin{eqnarray}
\tau_{HS}=&=& d(\theta,\zeta,\phi) \sin\zeta \sin\theta +\tau_0 \nonumber \\
\eta_{HS}&=&\frac{d(\theta,\zeta,\phi)}{\Lambda} \cos\theta \nonumber \\
x_{HS}&=& d(\theta,\zeta,\phi) \cos \zeta \sin\theta \cos\phi \nonumber \\
y_{HS}&=& d(\theta,\zeta,\phi) \cos \zeta \sin\theta \sin\phi \ ;\nonumber \\
&& 0\le \theta \le \pi \nonumber \\
&& 0\le \zeta \le \pi/2 \nonumber \\
&& 0\le \phi < 2 \pi \ ,
\label{eq:hyppar}
\end{eqnarray}
$\Lambda$ is a constant   length.

Following the Cooper-Frye prescription \cite{Cooper:1974mv}, particle
 spectra are given by
\begin{equation}
E\frac{d^3N}{dp^3}=\int d\Sigma_\mu p^\mu f(p_\mu u^\mu) \ .
\label{eq:statem}
\end{equation}
$d\Sigma_\mu=\epsilon_{\mu\nu\alpha\beta} \partial_\theta x^\nu \partial_\zeta
 x^\alpha \partial_\phi x^\beta d\theta d\zeta d\phi$
 is the integration element on the freeze-out hypersurface and
$f$ is the equilibrium Bose or Fermi momentum distribution. 
The four momentum of the emitted particle is
\begin{equation}
p^\mu= (m_\perp y, p_\perp \cos \phi_p,p_\perp \sin\phi_p,m_\perp y) \ ,
\end{equation}
and
\begin{equation}
p_\mu u^\mu=m_\perp \gamma \cosh(Y-y) -p_\perp(u_x \cos\phi_p+u_y \sin\phi_p) \ .
\label{eq:pu}
\end{equation}
\begin{eqnarray}&&
 d\Sigma_\mu p^\mu=\frac{1}{\Lambda}d^2 \sin \theta \left(\cos \zeta d^2 \sin \zeta \right. \nonumber \\&&
\left({p_\perp} \cos \zeta\cos(\phi-\phi_p)+m_\perp \cosh(y-\eta_{HS}) \sin \zeta\right) \sin^3 \theta\nonumber \\&&
-\cos \zeta \sin \theta ({p_\perp} {\tau_0} \cos \zeta \cos \theta\cos(\phi-\phi_p)\nonumber \\&&
+m_\perp \left({\tau_0} \cos \theta \cosh(y-\eta_{HS}) \sin \zeta\right.\nonumber \\&&
-\left.\left.\Lambda \sin \theta\sinh(y-\eta_{HS})\right)\right)\partial_\theta d\nonumber \\&&+
{\tau_0} \left(\cos \zeta \left(-m_\perp \cos \zeta \cosh(y-\eta_{HS})+{p_\perp}\cos(\phi-\phi_p) \sin \zeta\right) \right.\nonumber \\&&
\left.\partial_\zeta d+{p_\perp}\sin(\phi-\phi_p) \partial_\phi d\right)+d \sin \theta \nonumber \\&&
\left(\cos \zeta \left({p_\perp} {\tau_0} \cos \zeta\cos(\phi-\phi_p) \sin \theta+m_\perp \left({\tau_0} \cosh(y-\eta_{HS}) \right.\right.\right.\nonumber \\&&
\left.\left.\sin \zeta \sin \theta+\Lambda \cos \theta\sinh(y-\eta_{HS})\right)\right)\nonumber \\&&+
\sin \zeta \left(-\cos \zeta \cos \theta \left({p_\perp} \cos \zeta\cos(\phi-\phi_p)+m_\perp \right.\right.\nonumber \\&&
\left.\cosh(y-\eta_{HS}) \sin \zeta\right) \sin \theta \partial_\theta d\nonumber \\&&+
\cos \zeta \left(-m_\perp \cos \zeta \cosh(y-\eta_{HS})+{p_\perp}\cos(\phi-\phi_p) \sin \zeta\right) \nonumber \\&&
\left.\left.\left.\partial_\zeta d+{p_\perp}\sin(\phi-\phi_p) \partial_\phi d\right)\right)\right)d\theta d\zeta d\phi
\label{eq:dsigp}
\end{eqnarray}
After the hydrodynamic evolution,
the $3$ dimensional hypersurface parameterized by the variables 
$\theta, \zeta,\phi$ is exported to the statistical emission and
 resonance decay code THERMINATOR
\cite{Kisiel:2005hn}. The density (\ref{eq:statem}) (with (\ref{eq:pu})
 and (\ref{eq:dsigp})) is implemented in the code.
THERMINATOR generates events in two steps. First $380$ 
different kind of particles and resonances
emitted from the hypersurface are generated according to the density 
(\ref{eq:statem}), then resonances are allowed to decay.

\section{Particle spectra, flow, correlation radii}
\label{sec:spectra}

\begin{figure}
\includegraphics[width=.5\textwidth]{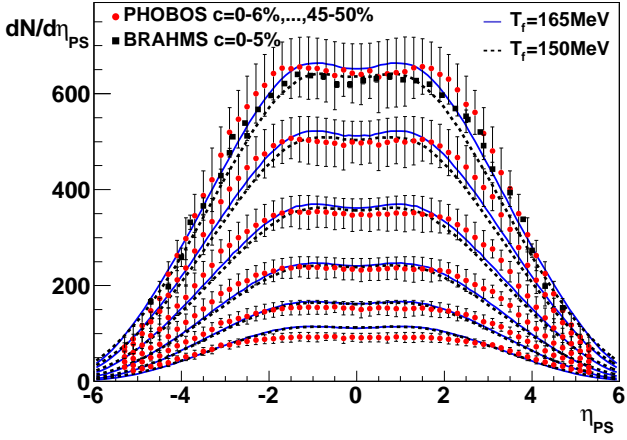}
\caption{(Color online) Pseudorapidity distribution 
of charged particles for centrality classes $0-6\%$, $6-15\%$, $15-25\%$, 
$25-35\%$, $35-45\%$ and $45-50\%$ calculated for the freeze-out
 temperatures  $T_f=165$ and $150$MeV (solid and dashed 
lines respectively) compared to PHOBOS Collab. data (dots) 
\cite{Back:2002wb}. The squares represent the BRAHMS Collab. 
data for centrality $0-5\%$ 
  \cite{Bearden:2001qq}.}
\label{fig:dndeta}
\end{figure}

\begin{figure}
\includegraphics[width=.4\textwidth]{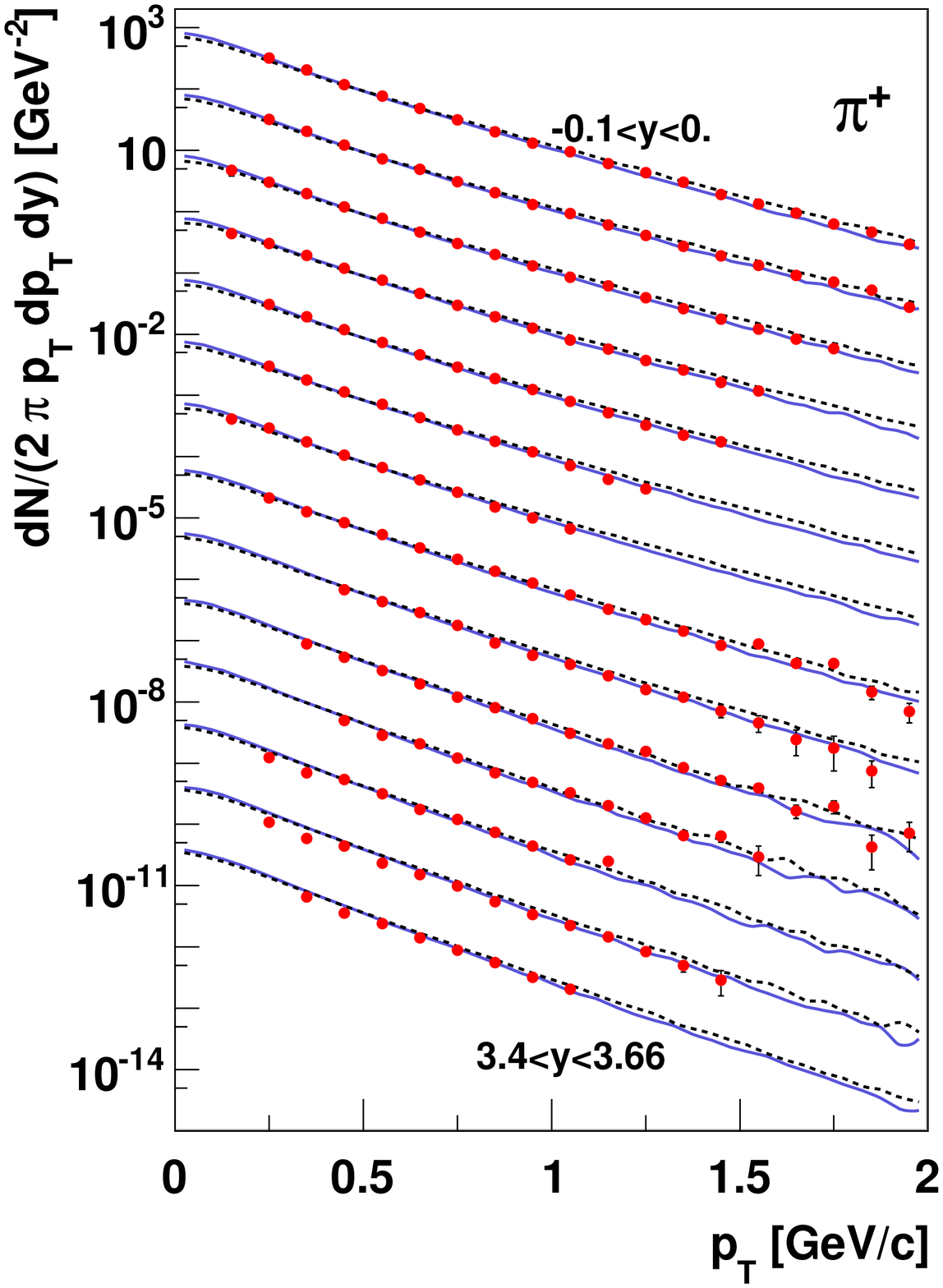}
\caption{(Color online) Transverse momentum spectra of $\pi^{+}$ for
 different rapidity windows 
ranging from $-0.1<y<0$ to $3.4<y<3.66$
 for centrality $0-5\%$ (results for different rapidity bins  are scaled down by 
powers of $1/10$). 
 The dots  represent the data of the BRAHMS Collab. 
  \cite{Bearden:2004yx}.}
\label{fig:ptrappiony}
\end{figure}

\begin{figure}
\includegraphics[width=.4\textwidth]{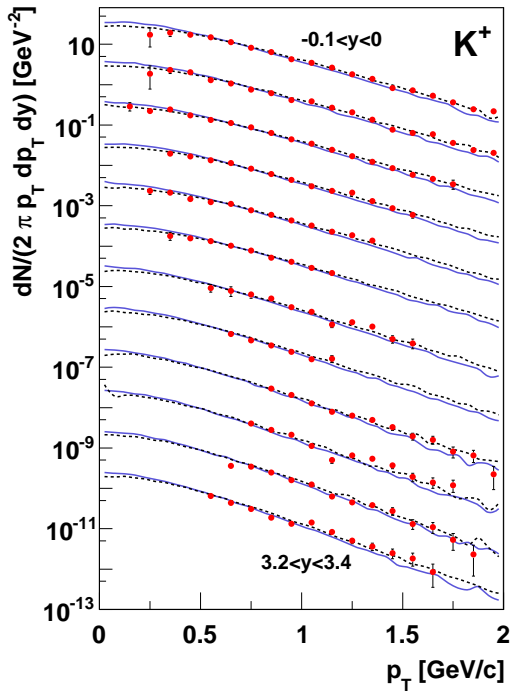}
\caption{(Color online) Transverse momentum spectra of $K^{+}$ for
 different rapidity windows 
ranging from $-0.1<y<0$ to $3.2<y<3.4$ for centrality $0-5\%$ 
(results for different rapidity bins are  scaled down by 
powers of $1/10$). 
 The dots  represent the 
data  of the BRAHMS Collab.
  \cite{Bearden:2004yx}.}
\label{fig:ptrapkaony}
\end{figure}

Charged particle distributions in pseudorapidity
 $\eta_{PS}=\frac{1}{2}\log\left( \frac{p+p_z}{p-p_z}\right)$
 have been measured 
for different centralities.
The $3+1$D hydrodynamic model can reproduce the data for centralities $0-40\%$
(Fig \ref{fig:dndeta}). 
We show  results for two freeze-out temperatures $T_f=165$ and $150$MeV.
The first one is close to the estimate of the chemical freeze-out 
temperature in Au-Au collisions
 \cite{BraunMunzinger:2001ip,Florkowski:2001fp,Cleymans:2004pp}.
When decreasing the freeze-out temperature particle multiplicity goes down,
but the effect is small. It gives confidence to our model, that assumes
 a chemically equilibrated fluid down to $T_f=150$MeV. For lower freeze-out 
temperatures the difference between a chemically equilibrated and a partially 
equilibrated fluid becomes significant \cite{Hirano:2002ds}.
The centrality dependence that we predict comes solely 
from the geometrical scaling of the fireball density according to Eq. 
(\ref{eq:nwdens}), other parameters (in particular the freeze-out temperature)
remain unchanged. On general grounds, one expects the 
hydrodynamic model to break down for very peripheral collisions.
The interaction region in peripheral collisions is not dense enough 
to equilibrate completely. Experimental data on the centrality dependence 
of  strangeness production and of  particle spectra suggest 
that at impact parameters $b>9$fm 
\cite{Bozek:2005eu,Bozek:2008zw,Becattini:2008ya} less than $70\%$ 
of the interaction region can be treated as a thermally equilibrated fireball.

\begin{figure}
\includegraphics[width=.4\textwidth]{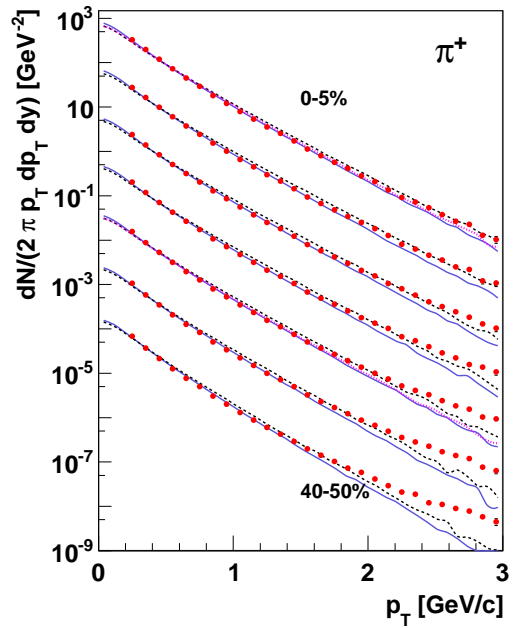}
\caption{(Color online) Transverse momentum spectra of $\pi^{+}$ for
 different  centrality classes $0-5\%$, $5-10\%$, $10-15\%$, 
$15-20\%$, $20-30\%$, $30-40\%$ and $40-50\%$ (results for different 
centralities
  are  scaled down by 
powers of $1/10$)  calculated for the freeze-out
 temperatures  $T_f=165$ and $150$MeV (solid and dashed 
lines respectively) compared to PHENIX Collab. data (dots) 
\cite{Adler:2003cb}. }
\label{fig:ptpiony}
\end{figure}

\begin{figure}
\includegraphics[width=.4\textwidth]{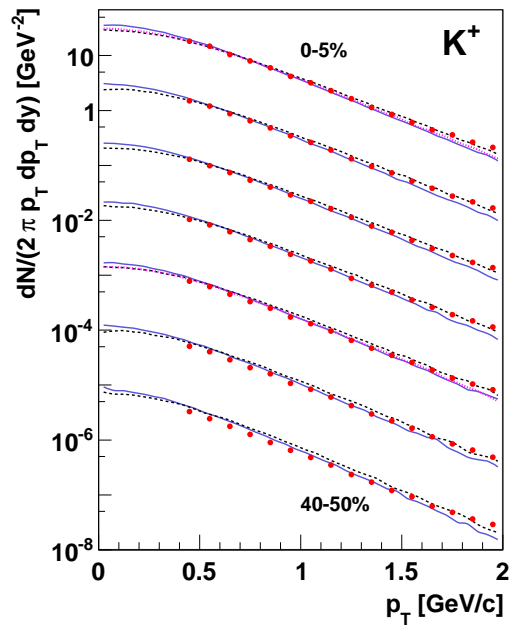}
\caption{(Color online) Same as Fig \ref{fig:ptpiony} but for $K^{+}$.}
\label{fig:ptkaony}
\end{figure}

For central collisions $c=0-5\%$ we  calculate  the transverse 
momentum spectra of pions and kaons at different rapidities.
In the whole rapidity range where identified particle spectra are 
available \cite{Bearden:2004yx}, we find an excellent agreement between
 the results of the hydrodynamic evolution coupled with statistical
 emission and the 
BRAHMS Collab. data (Figs. \ref{fig:ptrappiony} and \ref{fig:ptrapkaony}). 
The best agreement is achieved for a freeze-out
 temperature of $150$MeV. This very good agreement between the data 
and the hydrodynamic model indicates that
 the matter created in central collisions 
behaves as a thermally equilibrated  (in the rapidity range $-3.5<y<3.5$)
although not boost-invariant
 fireball.

A different way of testing thermalization in Au-Au collisions 
is to compare predictions and experimental data for transverse
 momentum spectra
 at different centralities. In Fig. \ref{fig:ptpiony} 
are shown $\pi^{+}$ spectra 
for $p_\perp$ up to $3$GeV/c
 and centralities in the range $0-50\%$. We 
find that  hydrodynamic calculations with $T_f=150$MeV are in 
 very good agreement with the experiment for all 
centralities and transverse momenta $p_\perp<2$GeV/c. Pions with higher 
transverse momenta originate mostly from hard processes and cannot be described
as emitted thermally from a collectively expanding fluid.
The  agreement between the experimental spectra and the results 
of the calculation for $K^{+}$ (Fig. \ref{fig:ptkaony}) is limited to 
centralities $0-30\%$. For centrality bins  $30-40\%$ and $40-50\%$ 
the calculation overpredicts the kaon multiplicity, even though the 
slope of the spectra is similar in the model and in the data.
The reduced strangeness production in peripheral collisions can be an 
effect of energy and momentum conservation \cite{Chajecki:2008yi}, 
canonical suppression \cite{Hamieh:2000tk}, or reduced size 
of the thermally equilibrated  fireball \cite{Bozek:2005eu,Becattini:2008ya}.
In Figs. \ref{fig:ptpiony} and \ref{fig:ptkaony} are also shown the results of 
a $2+1$D hydrodynamic calculation for two centralities $0-5\%$ 
and $20-30\%$ (dotted lines, indistinguishable from the $3+1$D results).
 The resulting spectra for central
 rapidities are very similar to the ones from the $3+1$D calculations, 
but  are obtained after the expansion of the  matter with smaller 
initial energy density.

\begin{figure}
\includegraphics[width=.35\textwidth]{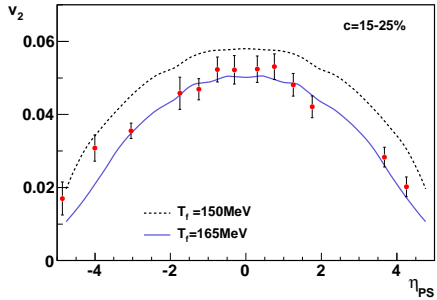}
\caption{(Color online) Pseudorapidity dependence of the 
elliptic flow coefficient for charged particles 
for centralities  $15-25\%$ for freeze-out temperatures $T_f=150$ and
 $165$MeV (dashed and solid lines respectively), data for the PHOBOS Collab.
 are denoted by dots
\cite{Back:2004mh}. }
\label{fig:v2rap}
\end{figure}

\begin{figure}
\includegraphics[width=.35\textwidth]{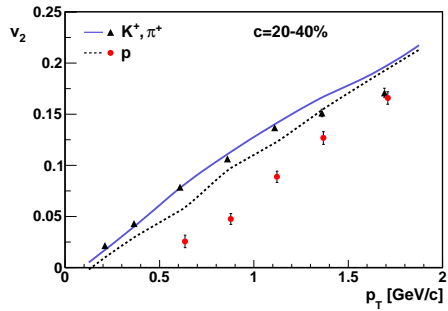}
\caption{(Color online) Transverse momentum dependence of the 
elliptic flow coefficient for protons (dotted line
 and dots) and for $\pi^{+}$
 and $K^{+}$ (solid line and triangles). Calculations are performed for 
$T_f=150$MeV, data are from the PHENIX Collab. 
\cite{Adler:2003kt}. }
\label{fig:v2pt}
\end{figure}

The elliptic flow represents a very sensitive probe
 of the collective behavior of the dense matter 
\cite{Voloshin:2008dg}.  The azimuthal
 asymmetry with respect to the reaction plane is described by the elliptic 
flow coefficient $v_2$
\begin{equation}
\frac{dN}{p_\perp d p_\perp d\phi_p }=\frac{dN}{p_\perp d p_\perp }
(1+2v_2(p_\perp)\cos(2\phi_p)+\dots) \ .
\end{equation}
The elliptic flow coefficient for charged particles 
has been measured for a broad range of pseudorapidities \cite{Back:2004mh},
showing a strong pseudorapidity dependence. There is no sign of a 
Bjorken plateau 
for central rapidities. To reproduce the shape of the $v_2$ pseudorapidity 
dependence, 
initial conditions in energy density with  a relatively narrow 
plateau in space-time 
rapidity must be chosen (Eq. \ref{eq:etaprofile}) \cite{Hirano:2005xf,Nonaka:2006yn}. 
Such initial conditions, combined with a hard equation of state and an early
initial time of the evolution result in a complete disappearance
 of the Bjorken plateau in the final hadron distributions. It must 
be stressed however, that it does not mean that for non-central rapidities
the evolution is not described by hydrodynamics and statistical emission.
The model model works very well and describes the observed spectra for
 $-3.5<y<3.5$ (Figs. \ref{fig:ptrappiony} and \ref{fig:ptrapkaony}).
The longitudinal expansion and the smaller size of the system at non-zero 
space-time rapidities reduce the final elliptic flow.
The elliptic flow as function of $p_\perp$ for identified particles is 
shown in Fig. \ref{fig:v2pt}. Hydrodynamic calculations describe
 the elliptic flow for mesons with $p_\perp<1.5$GeV/c, but to reproduce 
the saturation of $v_2$ for large $p_\perp$ dissipative or viscosity 
effects must be invoked. The elliptic flow for baryons is overpredicted 
for the freeze-out temperatures chosen.

\begin{figure}
\includegraphics[width=.35\textwidth]{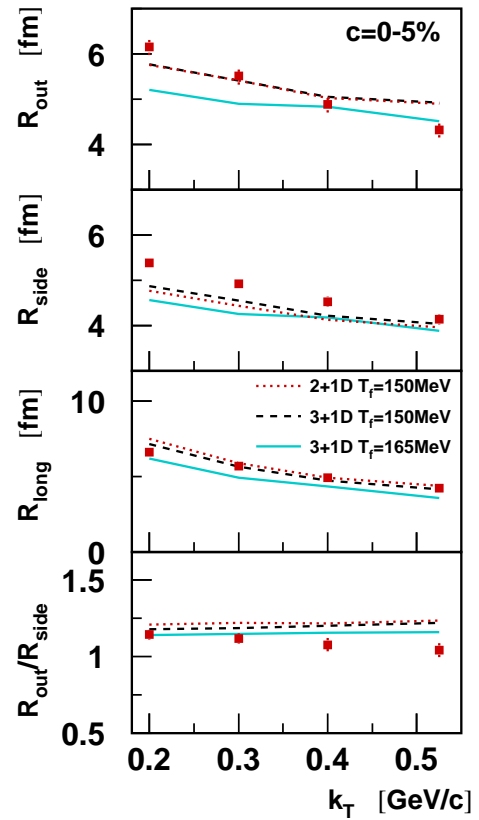}
\caption{(Color online) HBT radii for Au-Au collisions at centrality $0-5\%$.
$3+1$D calculations with $T_f=165$MeV (solid line) and  $T_f=150$MeV 
(dashed line), $2+1$D calculation with $T_f=150$MeV 
(dotted line) and STAR Collab. data \cite{Adams:2004yc} (squares) are shown. }
\label{fig:hbt}
\end{figure}

Pairs of identical particles emitted from the thermal source 
can be used to extract the size of the fireball from the interferometry 
measurement
\cite{Heinz:1999rw,Wiedemann:1999qn,Weiner:1999th,Lisa:2005dd}.
The statistical emission code 
THERMINATOR \cite{Kisiel:2005hn} provides the space-time points of  
particle creation. For a set of generated events the  correlation function is 
constructed from same-event and mixed-event  pairs
\cite{Kisiel:2006is}. Such a general procedure allows for an 
easy implementation of experimental cuts, final interaction or 
Coulomb corrections 
\cite{Bowler:1991vx,Sinyukov:1998fc} and can be also applied 
 to non-identical particle correlations
\cite{Kisiel:2006yv}.
The extracted multidimensional correlation functions 
are parameterized by the Bertsch-Pratt 
formula
\cite{Pratt:1986cc,Bertsch:1988db}
\begin{eqnarray}
C(k_\perp,q_{out},q_{side},q_{long})&=&1+\lambda
 \exp(-R_{side}^2 q_{side}^2\nonumber \\
&-&R_{out}^2 q_{out}^2-R_{long}^2 q_{long}^2) \ .
\end{eqnarray}
The fit parameters $R$ (HBT radii) are extracted for fixed bins
 of pair total transverse momenta $k_\perp$. The experimental radii $R_{out}$, 
$R_{side}$ 
and $R_{long}$ are well described by the hydrodynamic calculations 
with $T_f=150$MeV (Fig. \ref{fig:hbt}). We notice that the $3+1$D and $2+1$D 
calculations that both describe the transverse momentum 
spectra, give also very similar HBT radii. Similarity in the spectra 
means that the transverse collective flow is the same; together with the 
similarity in  the freeze-out hypersurfaces (Fig. \ref{fig:freeze1}) it 
explains why 
 the HBT radii from the $3+1$D and $2+1$D evolutions come out so close. 
 The discrepancies between the calculations and the data are smaller than
 $10\%$. The ratio $R_{out}/R_{side}$ from the 
model comes out  close to the data as well. This remarkable property of
 modern hydrodynamic calculations, 
which solves the so called RHIC HBT  puzzle, has been noticed in Ref. 
\cite{Broniowski:2008vp} based on $2+1$D simulations.
The present study is the first $3+1$D calculation using the
 same equation of state and an early initial time.

\section{Conclusions}

We present an extensive study of the $3+1$D hydrodynamic model
 of the evolution of the fireball. Compared to other similar calculations we
 use a different equation of state \cite{Chojnacki:2007jc} and an
 early initial time to start up  the expansion. As a result we find 
a very good agreement between the calculated transverse momentum spectra 
for pions and kaons for different centralities and a broad range of rapidities.
The fireball formed in Au-Au collisions at $\sqrt{s}=200$GeV is thermalized 
for centralities  $0-40\%$. The result is  remarkable, 
since we use an energy density profile for different impact parameters
 predicted from the Glauber Model geometrical scaling, with an overall
 normalization fixed for most central collisions. It is not surprising 
that for peripheral collisions this scaling breaks down \cite{Bozek:2008zw}, and
we overpredict the size of the thermalized fireball. The 
thermal description of the 
 production of 
strange particles is subject to even more restrictions. As a
 result the spectra of kaons are well reproduced for centralities $0-30\%$. 
For more peripheral collisions  the number of observed kaons is smaller than 
predicted in the model, signaling an incomplete chemical equilibration of the
interaction region.

The caculation reproduces the observed spectra up to  transverse
 momenta of $2$GeV/c, in a wider range  than most of the previous 
calculations. This indicates that  a
 long hydrodynamical evolution generates
the 
correct  amount of collective flow.  For central collisions we
 obtain an excellent description of 
pion and kaon spectra for non-central rapidities.
This demonstrates that the fireball is thermalized and that  particle 
production is statistical for all rapidities in the range $\pm 3.5$ units. 
This confirms in a dynamical calculation the applicability of the 
statistical fits from Ref. 
\cite{Biedron:2006vf}. The elliptic flow shows a strong pseudorapidity 
dependence, that can be reproduced assuming a collective thermal evolution 
but with reduced initial energy density at non-zero  space-time rapidities.

Another important result is the satisfactory description of pion
 interferometry radii. Assuming a rapid expansion of the system, the right
 amount of transverse flow and a 
reasonable life-time of the fireball are obtained.
This gives HBT radii $R_{out}$, $R_{side}$ and $R_{long}$ similar
 as in the experiment. The ratios $R_{out}/R_{side}$ come out 
to within less than  $10\%$ of the measured values. We also show that
 very similar HBT radii can be obtained from a $2+1$D hydrodynamic 
calculation with correctly chosen initial conditions.
It must be stressed however, that the initial energy density that reproduces
 the experimental spectra and HBT radii is 
$86$GeV/fm$^{3}$ at the maximum in $2+1$D, whereas it is $107$GeV/fm$^{3}$
 in $3+1$D. As mentioned, our $3+1$D calculation  describes 
particle emission at non-zero rapidities as well.

\section*{Acknowledgments}
Discussions with  Miko\l aj Chojnacki, Wojtek Florkowski and
 Tetsufumi Hirano  are gratefully acknowledged. The authors thank  Adam Kisiel
for providing the pair correlations analysis code.
\bibliography{../hydr}

\end{document}